\documentstyle[prb,aps]{revtex}
\input{epsf}
\draft
\begin{document}

\title{Applicability of the Broken-Bond Rule to the Surface Energy
of the fcc Metals}

\author{I. Galanakis$^1$, N. Papanikolaou$^2$, and P. H. Dederichs$^1$}

\address{$^1$Institut f\"ur Festk\"orperforschung, Forschungszentrum J\"ulich, D-52425 
J\"ulich, Germany\\
$^2$Fachbereich Physik, Martin-Luther Universit\"at, Halle-Wittenberg, 
D-06099 Halle, Germany} 

\date{\today}
\maketitle

\begin{abstract}

We apply the  Green's function based full-potential screened
Korringa-Kohn-Rostoker method in conjunction with the local
density approximation to study the surface energies of the noble
and the fcc transition and $sp$ metals.  The orientation
dependence of the transition metal surface energies can be well
described taking into account only the broken bonds between first
neighbors, quite analogous to the behavior we recently found for
the noble metals [see cond-mat/0105207]. The (111) and (100) surfaces of the $sp$ metals
show a jellium like behavior but for the more open surfaces we
find again the noble metals behavior but with larger deviation
from the broken-bond rule compared to the transition metals. 
Finally we show that the
use of the full potential is crucial to obtain accurate surface
energy anisotropy ratios for the vicinal surfaces.
\end{abstract}


\section{INTRODUCTION}

The surface energy is one of the most fundamental solid state
properties since it determines the equilibrium shape of a
mezoscopic crystal and plays a decisive role in phenomena like
roughening, faceting and crystal growth. Despite of their
importance, surface energies are difficult to determine
experimentally and just few data exist \cite{kumikov}. Most of
these experiments are performed at high temperatures and contain
uncertainties of unknown magnitude \cite{kumikov}. The most
comprehensive experimental data stem from surface tension
measurements in the liquid form being extrapolated to zero
temperature \cite{Tyson,Boer}, which cannot provide orientation
specific information. Gold crystallites \cite{gold-cryst} and
single surfaces \cite{breuer} have attracted a lot of  attention
aiming to study the orientation dependence of the surface energy,
but these experiments, as it was
 also the case for experiments on
In and Pb crystallites  \cite{Metois}, are performed at high
temperatures so that the results are difficult to interpret.
Entropy terms, describing the lower vibrational frequencies of the
atoms at the surface as compared to the bulk, the formation of
kinks and finally the creation of holes and pillboxes at the
low-index surfaces, have to be added to the total free energy. At
such high temperatures
 the  surface-melting faceting \cite{frenken}, i.e.  the break-down
of a vicinal surface in a dry and a
melted one, plays a predominant role.
 Also the measurement of core level shifts at the
surface has been proposed as an indirect measurement of the
surface energy anisotropy \cite{bonzel3}.  Recently, Bonzel and
Edmundts \cite{bonzel2}
 have shown that analyzing the equilibrium
shape of crystallites at various temperatures by
scanning tunneling microscopy can yield absolute values of
 the surface energies versus temperature,
but this technique has not yet been applied.

During the last years there have been several attempts to
calculate the surface energy of metals using either {\em
ab-initio} techniques \cite{Methfessel,SkriverPRB92,Kollar},
tight-binding parameterizations \cite{tb1,tb2} or semi-empirical
methods \cite{semi,rodriguez}. Methfessel and collaborators were
the first to study the trends in the surface energy, work function
and relaxation for the whole series of bcc and fcc 4$d$ transition
metals \cite{Methfessel}, using a full-potential version of the
linear muffin-tin orbitals (LMTO) method in conjunction with the
local-spin density approximation to the exchange-correlation
potential \cite{kohn,barth}. In the same spirit Skriver and
co-workers have used a Green's function LMTO technique
\cite{SkriverPRB91} to calculate the surface energy and the work
function of most of the elemental metals
\cite{SkriverPRB92,Kollar,Alden}. Recently, Vitos and 
collaborators \cite{VitosPRB97} using their full-charge density (FCD)
Green's function LMTO technique   in the atomic sphere
approximation (ASA) \cite{ASA} in conjunction with the generalized
gradient approximation (GGA) \cite{GGA} constructed a large
database that contains the low-index surface energies for 60
metals in the periodic table \cite{VitosSurf98}. Their results
present a mean deviation of 10 \% from the full-potential results
by Methfessel and collaborators for the 4$d$ transition metals
\cite{VitosPRB94}. Afterwards, they have used this database in
conjunction with the pair-potential model \cite{Moriarty} to
estimate the formation energy for monoatomic steps on low-index
surfaces for an ensemble of the bcc and fcc metals
\cite{VitosSurf99}.

In reference \onlinecite{PRL} we have demonstrated that the surface
energies of noble metals scale accurately with the number of
broken bonds between first neighbors. This broken-bond rule is
very useful for the estimation of the surface energies of vicinal
surfaces and  of the step energies; the latter ones can be
calculated as the energy difference between a vicinal and a flat
surface. In this contribution we investigate the question whether
the broken-bond rule can also be applied to the surface energies
of the other paramagnetic fcc metals: the transition metals Rh,
Pd, Ir and  Pt, and the $sp$ metals Ca, Sr, Al and Pb. To
calculate the surface energies we used the   recently developed
screened Korringa-Kohn-Rostoker (KKR) method  which  has been
already used to calculate the magnetic properties of 4$d$
monoatomic rows on Ag vicinal surfaces \cite{valerio}. In Section
II we analyze the details of our calculations and the convergence
of our results. We also discuss the importance of relativistic
effects. In Section III, we present the surface
energies of the transition and $sp$ metals and discuss the
applicability of the broken bond rule for these systems.  All
results presented in section III  are obtained  accounting for 
relativistic effects in the scalar-relativistic approximation.
Finally we discuss the use of the  full-potential instead of the atomic
sphere approximation

\section{METHOD OF CALCULATION}

\subsection{Computational Details}

To perform the calculations, we used the Vosko, Wilk and Nusair
parameterization \cite{Vosko} for the local density approximation
(LDA) to the exchange-correlation potential \cite{kohn} to solve
the Kohn-Sham equations within the screened KKR method that was
recently developed in our group \cite{Zeller95}. Its main
advantage is that it can treat both 2$D$ and $3D$ systems in the
same footing. Both the atomic sphere approximation (ASA)
\cite{ASA} and the capability to treat the full-potential (FP) are
implemented in this scheme. The ASA calculations take into account
the full charge density. It was shown by Andersen and
collaborators that the charge density obtained in this way for
spherically symmetric potentials is close to the density obtained
using a FP method \cite{Andersen86}. The full-potential is
implemented by using a Voronoi construction  of Wigner-Seitz
polyhedra that fill the space as described in reference
\onlinecite{Stefanou}. A repulsive muffin-tin potential (4 Ry high) is
used as reference system to screen the free-space long-range
structure constants into exponentially decaying ones
\cite{Zeller97}.  For the screening we took for all metals
interactions up to the second neighbors into account leading to a
tight-binding (TB) cluster around each atom of 19 neighbors. To
calculate the charge density, we integrated along a contour on the
complex energy plane, which extends from the bottom of the band up
to the Fermi level \cite{Zeller82}. Due to the smooth behavior of
the Green's functions for complex energies, only few energy points
are needed; in our calculations we used 27 energy points. For the
Brillouin zone (BZ) integration, special points are used as
proposed by Monkhorst and Pack \cite{monkhorst}. Only few tens of
${\bf k}_\parallel$ are needed to sample the BZ for the complex
energies, except for the energies close to the real axis near the
Fermi level for which a considerably larger number of ${\bf
k}_\parallel$ points is  needed. Here we used from $\sim$ 300
points for the vicinal surfaces up to $\sim$ 800 points for the
(110) surface. In addition we used a cut off of $\ell_{max}$=6 for
the mutlipole expansion of the charge density and the potential
and a cut off of $\ell_{max}$=3 for the wavefuctions. Finally in
our calculations the core electrons are allowed to relax during
the self-consistency.

To simulate the surface we used a slab with $N$ metal layers and
$N_{vac}$ vacuum layers  from each side. We have converged the
number of metal and vacuum layers so that our surface
energies are converged within 0.01 eV. The number of layers needed
to converge the surface energies increases with the  roughness of
the surface  and we had to use 12 layers of the fcc metal for the
(111) surface, 14 for the(100), 18 for the (110), 21 for the
(311), 30 for the (331) and the (210) surfaces. We have also used
3 vacuum layers from each side of the slab. For all the systems
studied we used the experimental lattice parameters: 3.80 \AA\ for
Rh, 3.89 \AA\ for Pd,  3.84 \AA\ for Ir, 3.92 \AA\ for Pt, 5.58
\AA\ for Ca, 6.08 \AA\ for Sr, 4.05 \AA\ for Al and finally 4.95
\AA\ for Pb \cite{ashcroft}. These numbers differ around 0.1 \AA\
from the numbers used in reference \onlinecite{VitosSurf98} where the
theoretical GGA equilibrium lattice constants have been used.

\subsection{Stability of Anisotropy Ratios}

To test our convergence we present in table \ref{table1} the
scalar-relativistic ASA  low-index surface energies of Ag and the
anisotropy ratios  with respect to the different parameters used
in the program. The absolute values of the energies change less
than 0.01 eV and the anisotropy ratios change by less than 1\%.
The largest effect comes from the $\ell_{max}$ cut-off for the
wavefunctions, but globally the first set of parameters is
sufficient to give accurate values of both the surface energies
and the anisotropy ratios. The second test presented in table
\ref{table2} concerns the effect of the lattice parameter on the
surface energies and on the anisotropy ratios. Our test has been
also performed for the noble metals in scalar-relativistic ASA.
Using Cu as a test case, the absolute values of the surface
energies change by the same percentage for all the surface
orientations when the lattice parameter is decreased. In the case
of Cu the theoretical LDA lattice constant is around 2\% smaller
than the experimental one, so that this effect changes the surface
energy  by less than 2.5\%. However due to error cancellation the
anisotropy ratio changed by less than 1.1\%. For Ag and in general
for the other 4$d$ metals, Rh and Pd, the LDA lattice constants
are somewhat 1\% smaller than the experimental lattice constants,
while for the 5$d$ metals the LDA values agree with the
experimental lattice constants. Therefore the anisotropy ratios
are independent of whether we use the LDA or the experimental
lattice parameter and  the following discussion of the anisotropy
ratios would not change if we would have used the LDA instead of
the experimental lattice constants in our calculations.

\subsection{Relativistic Effects}

\begin{figure}
\begin{center}
  \begin{minipage}{2.5in}
\epsfxsize=2.5in \epsfysize=2.5in \centerline{\epsfbox{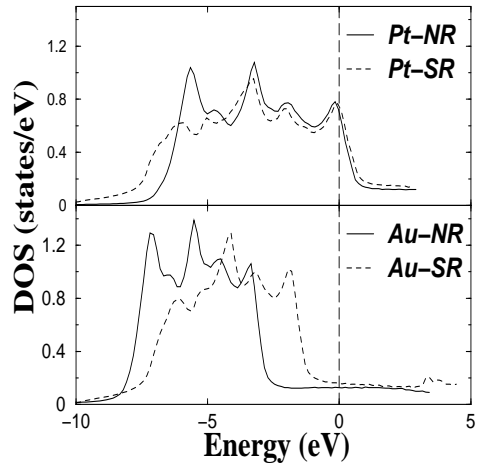}}
  \end{minipage}
\end{center}
   \caption{Non-relativistic (NR) density of states (DOS) of Pt (upper panel)
   and Au (bottom panel) compared with the
scalar-relativistic (SR) DOS. All DOSs are calculated for the
central layer atom of the (111) slab.
   \label{fig1}}
  \end{figure}

To study the importance of the relativistic effects we present in
figure \ref{fig1} the non-relativistic (NR) and scalar
relativistic (SR) density of states (DOS) for Au and Pt for the
central layer atoms of  the (111) slab which represent accurately
the bulk DOS. Au has all $d$ states filled and the $d$-bands are
deeper in energy compared to Pt that has only 9 $d$-electrons and
thus has a peak near the Fermi level.  This means that the bonds
between Pt atoms are stronger than for Au, so that for Pt more
energy is needed to create a  surface of a given orientation than
for Au. Relativistic effects are small for the Pt DOS at the Fermi
level but they widen the $d$-band so that the energy needed to
break a bond is considerably increased. Although the SR-DOS of Au
at the Fermi level is barely enhanced, relativistic effects
broaden the band and shift it higher in energy thus enhancing the
bonds between Au atoms and increasing the  surface energies. Ag
and Cu present a similar DOS with Au although the bandwidth of
their $d$-bands is somehow smaller. In these cases, relativity
only slightly changes the DOS, so the effect on the surface
energies is smaller than for Au.

\begin{figure}
\begin{center}
  \begin{minipage}{2.5in}
\epsfxsize=3.0in \epsfysize=2.5in \centerline{\epsfbox{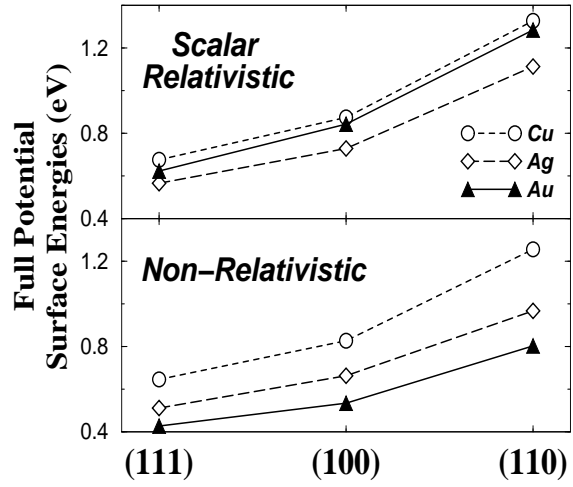}}
  \end{minipage} \end{center} 
\caption{Effect of relativity on the full-potential surface
energies of the noble metals. The effect of relativity is large
for Au that possesses 5$d$ electrons. Relativity also enhances the
anisotropy ratios.
 \label{fig2}}
 \end{figure}

In figure \ref{fig2} we present the FP  surface energies of the
three noble metals obtained both in NR and  SR calculations. The
main relativistic effect is  to increase the surface energies and
this effect is largest for Au being the heaviest element. In the
NR calculations Au has the lowest surface energies of all three
noble metals but in the SR calculations Au surface energies become
comparable to the one of Cu and the inclusion of spin-orbit
coupling further increases them \cite{PRL}. Also the anisotropy
ratios are increased compared to non-relativistic values, by about
 2-4\% in the case of Cu and Ag and up to 8-10\% in the case
of Au.

\section{SURFACE ENERGIES AND ANISOTROPY RATIOS}

We have studied the surface energies of the low-index surfaces,
(111), (100) and (110) and of the three most close-packed vicinal
surfaces: (311), (331) and (210). From a slab calculation one can
calculate the surface excess free energy at zero temperature from
the relation
\begin{equation}
\gamma=\frac{E_{slab}-N E_{bulk}}{2} \label{eq1}
\end{equation}
\noindent where $E_{slab}$ is the total energy of the slab, $N$ is
the total number of layers of the metal, $E_{bulk}$ is the per
atom energy in the bulk crystal and the 2 enters because when we
do a slab calculation we open two surfaces. To be consistent for
all the cases we used as $E_{bulk}$ the energy per atom of the
central layer of the slab.

The broken bond-rule states \cite{PRL}  that the surface energy
 $\gamma_{(hkl)}$ in eV/(surface atom) needed to create a surface with
a Miller index $(hkl)$ reduces just to the product  of
$\gamma_{(111)}$ and the ratio of the first-neighbor broken bonds
$N_{(hkl)}$ and $N_{(111)} = 3$:
\begin{equation}
 \gamma_{(hkl)} \cong \frac{N_{(hkl)}}{3} \gamma_{(111)}.
\end{equation}
$N_{(hkl)}$ can be easily obtained for any fcc surface~\cite{mackenzie}:
\begin{equation}
N_{(hkl)} = \left\{ \begin{array}{ll} 2h+k & \mbox{$h,k,l$ odd} \\
4h+2k& \mbox{otherwise} \end{array} \right. \qquad h \ge k \ge l.
\end{equation}

As a consequence the anisotropy ratios between the surface energy
for an arbitrary surface orientation over the surface energy for
the (111) surface orientation is close to the ratio between the
number of broken bonds between nearest neighbors for these
surfaces. Here we will study how well the paramagnetic fcc
transition and $sp$ metals satisfy the broken bond rule for the
surface energies.

\subsection{Transition Metals}

In table \ref{table3} we have gathered  our FP  scalar
relativistic results for all the transition metal surfaces. The
first comment that someone can make on this table is pretty
obvious. For all the metals the surface energy increases with the
roughness of the surface, i.e. as the number $N_{(hkl)}$ of broken
bonds increases, where (111) is the most close-packed surface with
$N_{(111)}$=3. Also it is pretty obvious that surface energies for
an isoelectronic row increase with the extent of the wavefunction;
the surface energy of Pt is larger than the one of Pd, since the
Pt 5$d$ wavefunctions have a somewhat larger extent than the Pd
4$d$ ones. Also along a line surface energy is larger for the
compound with the smaller number of $d$ electrons as this one
presents a stronger peak at the Fermi level (see figure
 \ref{fig1}), e.g. Pt presents larger surface energies than Au. Also in the same
table we present in parenthesis the anisotropy ratios. To open the
(100) surface we break 4 nearest-neighbor bonds, the (110) 6
bonds, the (311) 7 bonds, the (331) 9 bonds and finally the (210)
10 bonds. So the ideal broken bond ratios with respect to the
(111) surface, for which we break 3 nearest-neighbors bonds, are
4/3, 2, 7/3, 3 and 10/3, for the (100), (110), (311), (331) and
(210) surface orientations respectively. The calculated surface
energy anisotropy ratios deviate slightly from these ideal
numbers. For Pd the ratios are  smaller by the ideal ones by
$\sim$3-4\% for all the surface orientations, while Ir ratios are
larger from the ideals ones by $\sim$4-7\% for all the surface
orientations except the (331) where the Ir calculated ratio is
only by 1.3\% larger than the ideal ratio. Pt and Rh
 show a mixed behavior but in general the ratios differ less than 3\% from the ideal
nearest-neighbors broken bonds ratios for any surface. So in general
transition metal surfaces follow the broken bond rule but with slightly larger
deviations than the noble metals due to the fact that their $d$ band is not filled and
they present peaks at the Fermi level, which can slightly change
from one surface orientation to the other and consequently the energy needed to break
a bond changes also slightly.

\begin{figure} \begin{center}
  \begin{minipage}{2.5in}
\epsfxsize=3.0in \epsfysize=2.5in \centerline{\epsfbox{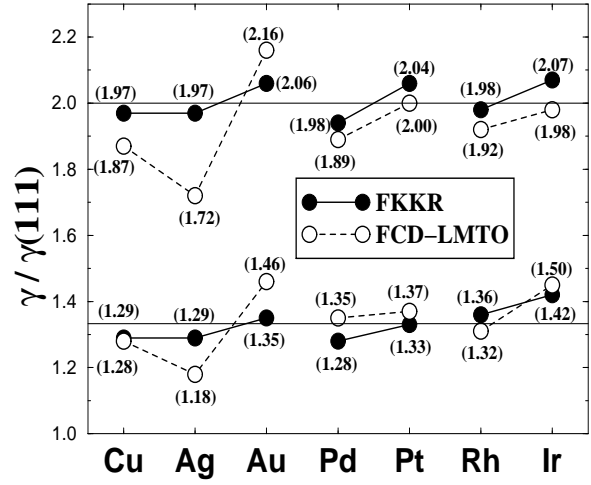}}
  \end{minipage}
\end{center} 
 \caption{ Anisotropy ratios within both FKKR and LMTO from reference 25.
 Contrary to the noble metals, whose (111)
surface posses a surface state centered at the $\bar\Gamma$ point,
the transition metals results are similar for both methods.}
   \label{fig3}
  \end{figure}

In figure \ref{fig3} we have plotted the surface energy anisotropy
ratios with respect to the (111) surface for the low-index surface
together with the results by Vitos {\em et al.} from the reference
\onlinecite{VitosSurf98}. In contrast to the noble metals the values of
Vitos {\em et al.} only slightly deviate from our results for the
transition metals.  In reference \onlinecite{PRL} we have explained the
discrepancy for the noble metals as due to an insufficient number
of {\bf k}$_\parallel$-points used in reference \onlinecite{VitosSurf98}
in the evaluation of the Brillouin zone integrals. The (111)
surface of the noble metals presents a surface state centered at
the $\bar{\Gamma}$ point which can only be accounted for by a
sufficiently dense  grid. Such states do not occur for the fcc
transition metals and thus the number of {\bf
k}$_\parallel$-points used in reference \onlinecite{VitosSurf98} is
sufficient to produce accurate ratios, that agree well with our FP
results. Since the authors of the reference  \onlinecite{VitosSurf98}
did not perform calculations for the vicinal surfaces, we cannot
judge the behavior of the FCD-LMTO-ASA method in these more
difficult cases.

As mentioned in the introduction there are several other {\em
ab-initio} calculations for the surface energies of these
materials but they concern just one or two surfaces and thus allow
no conclusions for the anisotropy ratios. In table \ref{table4} we
have gathered the results from previous {\em ab-initio}
calculations and experiments. We have expressed all the results in
Jm$^{-2}$ and not in eV/atom, as experiment have been performed in
the liquid phase of the metals. For the noble
 metals our and Vitos' results agree nicely when expressed in  Jm$^{-2}$, and the
calculated values are very close to the experimental ones. The
latter values are not orientation specific but averaged values and
they should be closer to the most close-packed surface: the (111).
Other calculations agree with our and Vitos databases. The FP-LMTO
results by Methfessel and collaborators \cite{Methfessel} agree
nicely with ours except the case
 of Ag where they predict a jellium like behavior for the (111) and (100) surfaces, i.e.
same surface energy per unit surface area, which is not expected
for a noble metal. Finally our calculated surface energies are in
reasonable agreement  with previous tight-binding calculations  by
Barreteau {\em et al.} \cite{tb1} on the low-index surfaces of Rh
and Pd and by Mehl and Papaconstantopoulos \cite{tb2} on the
ensemble of  noble and transition metals.

We should mention that in our calculations we did not relax the
positions of the layers but to a large extent this effect does not
affect the surface energies \cite{PRL}. Feibelman and
collaborators \cite{Feibelman90,Feibelman92} and Mansfield and
collaborators \cite{Mansfield} showed by first-principle
calculations that the effect of the relaxation on the calculated
surface energy of a particular facet should be around 2-5\%
depending on the roughness of the facet. Surface relaxations for vicinal surfaces
 have been studied
mainly using semi-empirical methods due to the complexity  arisen
by the simultaneous relaxation of a large number of layers \cite{Wan}. Rodriguez and
collaborators using such a semi-empirical method showed that surface
relaxations affect typically the anisotropic ratios by less than 2
\% \cite{rodriguez}, and so the neglect of relaxation should have
little effect on our results. 

\subsubsection{ASA versus FP Calculations}

The results presented in the previous section have been obtained taking
into account the non-spherical part of the potential, i.e. known
as the full-potential scheme.
The use of the FP instead of the ASA accounts in a more accurate
way for the charge distribution near the surface where due to the
lower symmetry the charge exhibits larger variations than in the
bulk.  The use of FP lowers the energies compared to ASA by about
15\% for all the surfaces under study. A similar behavior is also
found  for the vicinal surfaces. At first sight it seems that the
ASA is efficiently accurate to describe the  surface energies of
the materials under study. However,  the
 anisotropy ratios are more sensitive that the surface energies themselves.

\begin{figure}
\begin{center}
  \begin{minipage}{3.0in}
\epsfxsize=3.0in \epsfysize=2.5in \centerline{\epsfbox{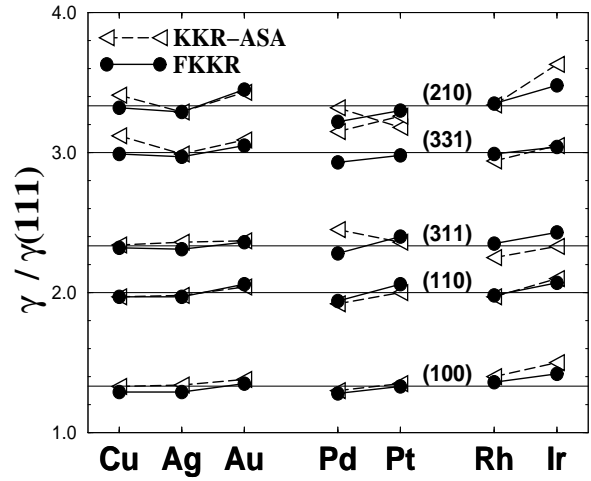}}
  \end{minipage}
\end{center} 
   \caption{ Anisotropy ratios for the low-index and the most close-packed
vicinal  surfaces
within ASA and FP. For the vicinal surfaces the use of FP is
necessary, due to the failure of ASA to describe accurately the
more complex metal-vacuum interface.}
   \label{fig4}
  \end{figure}

In figure \ref{fig4} we have represented the anisotropy ratios for
the noble and transition metals for the three low index and the
three vicinal surfaces which we have studied. The straight lines
represent the ideal broken bonds ratios. We see that already for
the Pd(311) surface the ASA produces a ratio that deviates
strongly from the broken bond rule while FP is near it. The
differences become even more dramatic for the more open (331) and
(210) surfaces. ASA produces ratios for the Cu and Ir surfaces
that deviate strongly from the broken bond rule while using the FP
we find again the ideal ratios. Especially for Pd, the ASA
predicts that the surface energy for the (331) surface is larger
than the one for the (210) surface which is more open. The use of
FP restores the broken-bond rule behavior. So although the ASA is
sufficiently accurate to produce reasonable surface energies, the
use of FP is decisive for the calculation of the anisotropy ratios
of the vicinal surfaces.

\subsection{$sp$ Metals}

\begin{figure}
\begin{center}
  \begin{minipage}{3.5in}
  \epsfxsize=3.5in \epsfysize=3.0in \centerline{\epsfbox{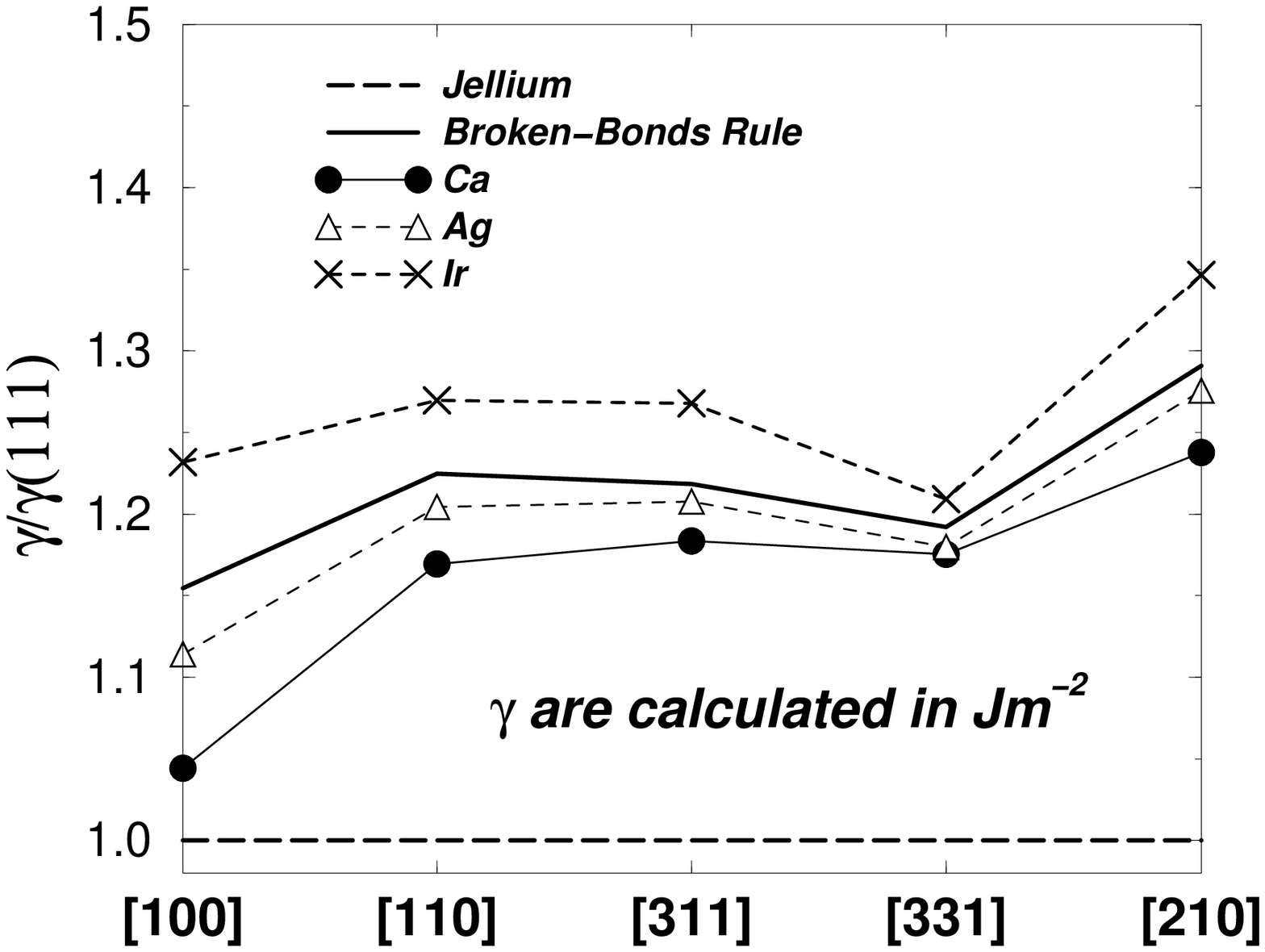}}
  \end{minipage}\end{center} 
  \caption{ Anisotropy ratios but calculating the surface energies in
Jm$^{-2}$ for Ag, Ir and Ca. The solid line corresponds to the
broken-bon rule and the tilted one to the jellium. The Ca(100)
surface is close to jellium but then it recovers the Ag and Ir
behavior.
   \label{fig5}}
  \end{figure}

To complete our study we also investigated the paramagnetic $sp$ metals -- Ca, Sr,
Al and Pb -- that crystallize also in the fcc structure.
In table \ref{table5} we have gathered
the FP surface energies and in parenthesis the anisotropy ratios for the six more
 close-packed surfaces. In general the surface energies for these
materials are smaller than for the $d$ metals due to the fact that
the bonds are made of  $s$ and $p$ electrons that are more mobile
than the localized $d$ electrons and so one needs less energy to
break these bonds. This becomes even more clear when we look at
the surface energy expressed in surface units (see table
\ref{table6}).   The Ca(100) and Sr(100) surfaces show  a rather
small anisotropy ratio compared to the transition and noble metals
and also compared to Al. Ca and Sr are in the periodic table just
near the simple metals  which are known to be well described by a
simple jellium model \cite{jellium}. So we expect that to some
extent we would find a jellium like behavior also for the Ca and
Sr surfaces at least for the low-index ones. This is really what
 happens. In figure \ref{fig5} we represent the anisotropy
ratios but now taking into account the energy per surface area
(Jm$^{-2}$) and not per atom (eV/atom). For a jellium model the
surface energy per surface area would be constant  for any surface
and the anisotropy ratio would be always 1. With the solid line we
represent the anisotropy ratios if the broken-bond rule is
applicable, and the ratios for Ca, Ag and Ir. Ag is closer to the
broken bond rule than Ir where the ratios are always
overestimated. For the Ca(100) surface we see that the value is
closer to the jellium model but for the more open surfaces the
anisotropy ratios are closer to the broken  bond model than the
jellium.

Contrary to Ca and Sr, Al shows the same behavior with Pd and the
calculated ratios are slightly smaller than  the ideal ones (see 
table~\ref{table5}). The most interesting case is Pb. For the (100)
and (110) surfaces the ratios are near the Sr ones while for the
next two vicinal surface orientations, (311) and (331), the
calculated anisotropy ratios are more than 6\% smaller than the
ideal values. But for the (210) surface the anisotropy ratio
changes the behavior and now is larger by 6\% than the ideal value
of 10/3. This is an indication that the behavior of Pb is
more complicated than all the other fcc metals we have studied in
this contribution. Here we have to mention that the surface
energies in table \ref{table5} have been calculated taking the
5$d$ as valence electrons as they are located just below the $sp$
bands. We recalculated the surface energies of Pb considering the
5$d$ as core states,  by increasing the number of layers, the
number of energy points used to do the integrations in the complex
energy plane and finally the number of {\bf k}$_\parallel$-points.
Although the surface energies changed slightly due to the core
treatment of the 5$d$ states,  the anisotropy ratios were
extremely stable.

In table~\ref{table6} we have gathered our calculated surface
energies in Jm$^{-2}$  together with  other calculations and
experiments. Our results agree nicely with the Vitos {\em et al.}
database  in reference \onlinecite{VitosSurf98} with the exception of Pb
where our surface energy per surface area is double as high, but
our results agree nicely with previous calculations by Mansfield
and Needs \cite{Mansfield} using pseudopotentials  and the
existing experimental data. Both Sch\"ochlin {\em et al.}
\cite{schochlin} and Stumpf and Scheffler \cite{stumpf} have
studied  all the three low-index surfaces of Al. Comparing
Sch\"ochlin {\em et al.} calculations with Stumpf's and
Scheffler's results we see that the former  calculations predict
comparable surface energies for the (100) and (110) surfaces while
the latter ones predict comparable surface energies for the (111)
and (100) surfaces. Both calculations are in contradiction to our
calculations that predict a considerable increase of the surface
energy as the surface becomes more open, while the calculations in
reference \onlinecite{VitosSurf98} state that the surface energy per
surface area is smaller for the (110) surface compared to the
(100) surface. This spread of the results for Al does not allow us
to draw safe conclusions for the variation of the surface energy
with the surface orientation.

\section{CONCLUSIONS}

We have shown using  the full-potential (FP) screened KKR code that the broken
bond rule, i.e. the surface energy scales linearly with the
number of nearest-neighbors broken bonds,  already shown for the noble metals
is also valid for the transition
metals.   For the $sp$ metals the (111) and (100)  surfaces show
a jellium like behavior but for the more open surfaces the surface energies follow
again the broken-bond rule with  Pb presenting the largest deviations.
The use of full-potential instead
of the full-charge atomic-sphere approximation (ASA) decreased the
surface energies and we showed that it is necessary to accurately
calculated the anisotropy ratios for the vicinal surfaces.

\section*{Acknowledgments}
Authors  gratefully acknowledge support from the TMR network
of {\em Interface Magnetism} (Contract No: ERBFMRXCT96-0089) and
the RT Network of {\em Computational Magnetoelectronics} (Contract
No: RTN1-1999-00145) of the European Commission.

\newpage

\begin{table}
\caption{Scalar-relativistic surface energies and anisotropy
ratios in parenthesis for Ag within ASA using different values for
the  the $\ell_{max}$ cut-off (LM), the tight-binding cluster
(TB), the number of energy points (EN) to perform integrations in
the complex energy plane and the number of {\bf
k}$_\parallel$-points in the full two-dimensional Brillouin zone. The first set of
parameters is the one used in the present calculations. }
\label{table1}
\begin{tabular}{ccccccrr}
LM & TB & EN & KP & Ag(111) & Ag(100)  & Ag(110) \\ \hline
3&19&27&55$\times$55 & 0.641 & {\em (1.34)} 0.860 & {\em (1.98)}
1.271 \\ 3&55&27&55$\times$55 & 0.637 & {\em (1.34)} 0.854 & {\em
(1.98)} 1.262 \\ 3&19&41&55$\times$55 & 0.643 & {\em (1.33)} 0.855
& {\em (1.97)} 1.266 \\ 3&19&27&75$\times$75 & 0.648 & {\em
(1.33)} 0.860 & {\em (1.96)} 1.273 \\ 3&55&41&75$\times$75 & 0.642
& {\em (1.32)} 0.849 & {\em (1.96)} 1.259 \\ 4&19&27&55$\times$55
& 0.649 & {\em (1.35)} 0.879 & {\em (2.00)} 1.301
\end{tabular}
\end{table}

\begin{table}
\caption{Effect of lattice parameter changes on the
scalar-relativistic ASA surface  energies
 for the three  low-index surfaces of Cu in the upper panel and for the (111) surface of
Cu, Ag and Au in the bottom panel,} \label{table2}
\begin{tabular}{ccccccr}
 & \multicolumn{2}{c}{Cu(111)} & \multicolumn{2}{c}{Cu(100)}
 & \multicolumn{2}{c}{Cu(110)}  \\
 $\frac{a_{exp}-a}{a_{exp}}$
& $\gamma$ (eV) & $\frac{\gamma_{exp}-\gamma}{\gamma_{exp}}$ &
$\gamma$ (eV) & $\frac{\gamma_{exp}-\gamma}{\gamma_{exp}}$ &
$\gamma$ (eV)& $\frac{\gamma_{exp}-\gamma}{\gamma_{exp}}$    \\
\hline 0\% & 0.737 &     & 0.982 &     & 1.455 &     \\ 1\% &
0.730 &0.95\%  & 0.971 & 1.12\% & 1.441 & 0.96\%  \\ 2\% & 0.719
&2.44\%  & 0.956 & 2.65\% & 1.420 & 2.41\%  \\ 3\% & 0.706 &4.21\%
& 0.939 & 4.38\% & 1.394 & 4.19\%  \\ \hline &
\multicolumn{2}{c}{Cu(111)} & \multicolumn{2}{c}{Ag(111)} &
\multicolumn{2}{c}{Au(111)}\\
 $\frac{a_{exp}-a}{a_{exp}}$
& $\gamma$ (eV) & $\frac{\gamma_{exp}-\gamma}{\gamma_{exp}}$ &
$\gamma$ (eV) & $\frac{\gamma_{exp}-\gamma}{\gamma_{exp}}$ &
$\gamma$ (eV)& $\frac{\gamma_{exp}-\gamma}{\gamma_{exp}}$    \\
\hline 0\% & 0.737 &     & 0.641 &     & 0.755 &     \\ 1\% &
0.730& 0.95\% &0.636 &0.78\% & 0.739 & 2.12\% \\ 2\% & 0.719&
2.44\% &0.624 &2.65\% & 0.717& 5.03\%\\ 3\% & 0.706 &4.21\% &0.609
&4.99\% & 0.692 & 8.34\%  \\
\end{tabular}
\end{table}

\begin{table}
\caption{Full-potential scalar-relativistic surface energies for
the six more closed-packed surfaces for the four transition
metals. In parenthesis the anisotropy ratios with respect to the
(111) surface.} \label{table3}
 \begin{tabular}{crrrr}
$\gamma$(eV) & Pd  & Pt  & Rh & Ir    \\  \hline (111) &
0.822 & 0.957 &1.034  & 1.200      \\ (100) & {\em (1.28)} 1.049 &
{(1.33)} 1.272 & {\em (1.36)} 1.404 & {\em (1.42)} 1.707  \\ (110)
& {\em (1.94)}  1.596 & {\em (2.06)} 1.973& {\em (1.98)} 2.047 &
{\em (2.07)} 2.488 \\ (113) & {\em (2.28)} 1.873&{\em (2.40)}
2.295 &  {\em (2.35)} 2.428  &{\em (2.43)} 2.913 \\ (331) &  {\em
(2.93)} 2.404  & {\em (2.98)} 2.853  & {\em (2.99)} 3.094 &
        {\em (3.04)} 3.652   \\
(210) &  {\em (3.22)} 2.644   & {\em (3.30)} 3.158  & {\em (3.35)}
3.464 &
        {\em (3.48)} 4.172
\end{tabular}
\end{table}

\begin{table}
\caption{Surface energies is J\thinspace m$^{-2}$ for the three
low-index surfaces of the paramagnetic fcc transition metals using the
FKKR. In second column the results using the FCD-LMTO-ASA and on
the third and fourth columns existing {\em ab-initio} calculations
and  experiments.} \label{table4}
\begin{tabular}{llccll}
\multicolumn{2}{c}{$\gamma$ (J\thinspace m$^{-2}$)} & FKKR &
 Ref. \onlinecite{VitosSurf98}
& Other Calc. & Experiments \\ \hline Cu & (111) & 1.91 & 1.95 &
1.59$^c$,1.94$^d$  & 1.79$^a$,1.83$^b$ \\
   & (100) & 2.15 & 2.17 & 1.71$^c$,1.80$^e$  & \\
   & (110) & 2.31 & 2.24 & 1.85$^c$      &  \\ \hline
Rh & (111) & 2.65 & 2.47 & 2.53$^f$,2.85$^g$     &
2.66$^a$,2.70$^b$ \\
   & (100) & 3.12 & 2.80 & 2.81$^f$,3.28$^g$,2.65$^h$,2.59$^i$ & \\
   & (110) & 3.22 & 2.90 & 2.88$^f$,3.37$^g$         & \\ \hline
Pd & (111) & 2.01 & 1.92 & 1.64$^f$           & 2.00$^a$,2.05$^b$
\\
   & (100) & 2.22 & 2.33 & 1.86$^f$,2.30$^j$,2.13$^k$ & \\
   & (110) & 2.39 & 2.23 & 1.97$^f$,2.50$^j$      & \\  \hline
Ag & (111) & 1.25 & 1.17 & 1.21$^f$           & 1.25$^a$,1.25$^b$
\\
   & (100) & 1.40 & 1.20 & 1.21$^f$,1.30$^j$,1.27$^l$,1.11$^m$ & \\
   & (110) & 1.51 & 1.24 & 1.26$^f$,1.40$^j$      & \\ \hline
Ir & (111) & 3.02 & 2.97 & 3.27$^n$      & 3.05$^a$,3.00$^b$ \\
   & (100) & 3.71 & 3.72 &                & \\
   & (110) & 3.82 & 3.61 &                & \\ \hline
Pt & (111) & 2.31 & 2.30 & 2.20$^n$,2.07$^o$  & 2.49$^a$,2.48$^b$
\\
   & (100) & 2.65 & 2.73 &                & \\
   & (110) & 2.91 & 2.82 &                & \\ \hline
Au & (111) & 1.39 & 1.28 & 1.25$^n$,1.04$^p$   & 1.51$^a$,1.50$^b$
\\
   & (100) & 1.62 & 1.63 & 1.33$^m$,1.30$^q$ &\\
   & (110) & 1.75 & 1.70 & 1.43$^r$&
\end{tabular} \ \\
{\small $^a$ Experiment, Ref. \onlinecite{Tyson}; $^b$ Experiment, Ref.
\onlinecite{Boer} \\ $^c$ Pseudopotentials, Ref. \onlinecite{rodach}; $^d$
FP-LMTO, Ref. \onlinecite{Polatoglou} \\ $^e$ modified APW, Ref.
\onlinecite{bross}; $^f$ FP-LMTO, Ref. \onlinecite{Methfessel}\\ $^g$
Pseudopotentials, Ref. \onlinecite{eichler};  $^h$ Pseudopotentials,
Ref. \onlinecite{morrison} \\ $^i$ FLAPW, Ref. \onlinecite{Feibelman90}; $^j$
FLAPW, Ref. \onlinecite{weinert}\\ $^k$ Pseudopotentials, Ref.
\onlinecite{wachter};  $^l$ FLAPW, Ref. \onlinecite{freeman}\\ $^m$
Pseudopotentials, Ref. \onlinecite{takeuchi2}; $^n$ Pseudopotentials,
Ref. \onlinecite{needs} \\ $^o$ Pseudopotentials, Ref.
\onlinecite{Feibelman95}; $^p$ Pseudopotentials, Ref. \onlinecite{takeuchi} \\
$^q$ FLAPW, Ref. \onlinecite{eibler}; $^r$ Pseudopotentials, Ref.
\onlinecite{Ho}
 }
\end{table}

\begin{table}
\caption{Full-potential scalar-relativistic surface energies for
the six more closed-packed surfaces of the four $sp$ metal. In
parenthesis the anisotropy ratios with respect to the
(111) surface.} \label{table5}
\begin{tabular}{crrrr}
$\gamma$(eV) & Ca& Sr& Al & Pb    \\  \hline (111) & 0.417 & 0.373
& 0.489 & 0.398 \\ (100) & {\em (1.21)} 0.503 & {\em (1.22)} 0.454
& {\em (1.28)} 0.625&{\em (1.24)} 0.492\\ (110) & {\em (1.92)}
0.797 & {\em (1.94)} 0.722 & {\em (1.93)} 0.943&{\em (1.95)}
0.778\\ (311) & {\em (2.27)} 0.946 & {\em (2.27)} 0.847 & {\em
(2.24)} 1.094&{\em (2.18)} 0.866\\ (331) & {\em (2.96)} 1.234 &
{\em (2.93)} 1.094 & {\em (2.94)} 1.436&{\em (2.81)} 1.118\\ (210)
& {\em (3.20)} 1.333 & {\em (3.18)} 1.187 & {\em (3.20)}
1.565&{\em (3.53)} 1.405
\end{tabular}
\end{table}

\begin{table}
\caption{Surface energies is J\thinspace m$^{-2}$ for the three
low-index surfaces of the paramagnetic fcc $sp$ metals using the
FKKR. In second column the results using the FCD-LMTO-ASA and on
the third and fourth columns existing {\em ab-initio} calculations
and  experiments.} \label{table6}
\begin{tabular}{llccll}
\multicolumn{2}{c}{$\gamma$ (J\thinspace m$^{-2}$)} & FKKR &
 Ref. \onlinecite{VitosSurf98}
& Other Calc. & Experiments \\ \hline Ca & (111) & 0.50 & 0.57 & &
0.50$^a$,0.49$^b$ \\
   & (100) & 0.52 & 0.54 & &\\
   & (110) & 0.58 & 0.58 & &\\ \hline
Sr & (111) & 0.37 & 0.43 &        & 0.42$^a$,0.41$^b$ \\
   & (100) & 0.39 & 0.41 & &\\
   & (110) & 0.44 & 0.43 & &\\ \hline
Al & (111) & 1.10 & 1.12 &0.94$^c$,1.12$^d$    & 1.14$^a$,1.12$^b$
\\
   & (100) & 1.22 & 1.35 &1.08$^c$,1.14$^d$ &\\
   & (110) & 1.30 & 1.27 &1.09$^c$,1.28$^d$ &\\ \hline
Pb & (111) & 0.60 & 0.32 &  0.50$^e$       & 0.59$^a$,0.60$^b$ \\
   & (100) & 0.64 & 0.38 & &\\
   & (110) & 0.72 & 0.45 &  0.59$^e$ &
\end{tabular} \ \\
{\small $^a$ Experiment, Ref. \onlinecite{Tyson}; $^b$ Experiment, Ref.
\onlinecite{Boer} \\ $^c$ Pseudopotentials, Ref. \onlinecite{schochlin}; $^d$
Pseudopotentials, Ref. \onlinecite{stumpf}\\ $^e$ Pseudopotentials, Ref.
\onlinecite{Mansfield} \\
 }
\end{table}

\newpage

\begin{thebibliography}{99}

\bibitem{kumikov}
V.K. Kumikov and Kh.B. Khokonov, {\em J. Appl. Phys.} {\bf 54}
(1983) 1346.

\bibitem{Tyson}
W.R. Tyson and W.A. Miller, {\em Surf. Sci.} {\bf 62} (1977) 267.

\bibitem{Boer}
F.R. Boer, R. Boom, W.C.M. Mattens, A.R. Miedema, and A.K.
Niessen, Cohesion in Metals, North-Holland, Amsterdam, 1988.

\bibitem{gold-cryst}
B.E. Sundquist, {\em Acta Met.} {\bf 12} (1964) 67; W.L.
Winterbottom and N.A. Gjostein, {\em Act. Met.} {\bf 14}  (1966)
1041; J.C. Heyraud and J.J. M\'etois, {\em Acta Met.} {\bf 28}
(1980) 1789; Z. Wang and P. Wynblatt,  {\em Surf. Sci.} {\bf 398}
(1998) 259.

\bibitem{breuer}
U. Breuer and H.P. Bonzel,  {\em Surf. Sci.} {\bf 273} (1992) 219.

\bibitem{Metois}
J.C. Heyraud and J.J. Metois, {\em Surf. Sci.} {\bf 128} (1983)
334; J.C. Heyraud and J.J. Metois, {\em Surf. Sci.} {\bf 177}
(1986) 213.

\bibitem{frenken}
G. Bilalbegivi\'c, F. Ercolessi, and E. Tosatti, {\em Surf. Sci.}
{\bf 280} (1993) 335; H.M. van Pixteren and J.W.M. Frenken, {\em
Europhys. Lett.} {\bf 21} (1993) 43; H.M. van Pixteren, B. Pluis,
and J.W.M. Frenken, {\em Phys. Rev. B} {\bf 49} (1994) 13798.

\bibitem{bonzel3}
H.P. Bonzel and K. D\"uckers,  {\em Surf. Sci.} {\bf 184} (1987)
425.

\bibitem{bonzel2}
H.P. Bonzel and A. Edmundts, {\em Phys. Rev. Lett.} {\bf 84}
(2000) 5804.

\bibitem{Methfessel}
M. Methfessel, D. Hennig, and M. Scheffler,  {\em Phys. Rev. B}
{\bf 46} (1992) 4816.

\bibitem{SkriverPRB92}
H.L. Skriver and N.M. Rosengaard, {\em Phys. Rev. B} {\bf 46}
(1992) 7157.

\bibitem{Kollar}
J. Koll\'ar, L. Vitos, and H.L. Skriver, {\em Phys. Rev. B} {\bf
49} (1994) 11288.

\bibitem{tb1}
C. Barreteau, D. Spanjaard, and M.C. Desjonqu\`eres, {\em Surf.
Sci.} {\bf 433-435} (1999) 751.

\bibitem{tb2} M.M. Mehl and D. Papaconstantopoulos, {\em Phys. Rev.
B} {\bf 54} (1996) 4519.

\bibitem{semi}  S.M. Foiles, M.I. Baskes, and M.S. Daw, {\em Phys. Rev. B} {\bf 33} (1996) 7983;
G.J. Ackland, G. Tichy, V. Vitek, and M.W. Finnis, {\em Phil. Mag.
A} {\bf 56} (1987) 735; D. Wolf,  {\em Surf. Sci.} {\bf 226}
(1990) 389; M.I. Baskes, {\em Phys. Rev. B} {\bf 46} (1992) 2727;
P. van Beurden and G.J. Kramer,  {\em Phys. Rev. B} {\bf  63}
(2001) 165106.

\bibitem{rodriguez}
A.M. Rodr\'iguez, G. Bozzolo, and J. Ferrante, {\em Surf. Sci.}
{\bf 289} (1993) 100.

\bibitem{kohn}
P, Hohenberg and W. Kohn,  {\em Phys. Rev.} {\bf 136} (1964) B864;
W. Kohn and L.J. Sham, {\em Phys. Rev.}  {\bf 140} (1965) A1133.

\bibitem{barth}
U. von Barth and L. Hedin, {\em J. Phys. C} {\bf 5} (1972) 1629.

\bibitem{SkriverPRB91}
H.L. Skriver and N.M. Rosengaard, {\em Phys. Rev. B} {\bf 43}
(1991) 9538.

\bibitem{Alden}
M. Alden, H.L. Skriver, S. Mirbt, and B. Johansson, {\em Phys.
Rev. Lett.} {\bf 69} (1992) 2296; M. Alden, H.L. Skriver, S.
Mirbt, and B. Johansson, {\em Sur. Sci.} {\bf 315} (1994) 157.

\bibitem{Feibelman95}
P.J. Feibelman, {\em Phys. Rev. B} {\bf 52} (1995) 16845.

\bibitem{VitosPRB97}
L. Vitos, J. Koll\'ar, and H.L. Skriver, {\em Phys. Rev. B} {\bf
55} (1997) 13521.

\bibitem{ASA}
O.K. Andersen, and O. Jepsen, {\em Phys. Rev. Lett.} {\bf
  53} (1984) 2671;
O.K. Andersen, O. Jepsen, M. Sob, in: M. Yussouf (Ed.), Electronic
Band Structure and its Applications, Springer-Berlin, 1987.

\bibitem{GGA}
J.P. Perdew, K. Burke, and M. Ernzerhof, {\em Phys. Rev. Lett.}
{\bf 77} (1996) 3865.

\bibitem{VitosSurf98}
L. Vitos, A.V. Ruban, H.L. Skriver and J. Koll\'ar, {\em Surf.
Sci.} {\bf 411} (1998) 186.

\bibitem{VitosPRB94}
L. Vitos, J. Koll\'ar, and H.L. Skriver, {\em Phys. Rev. B} {\bf
49} (1994) 16694.

\bibitem{Moriarty}
J.A. Moriarty and R. Phillips,  {\em Phys. Rev. Lett.} {\bf 66}
(1991) 3036.

\bibitem{VitosSurf99}
L. Vitos, H.L. Skriver and J. Koll\'ar, {\em Surf. Sci.} {\bf 425}
(1999) 212.

\bibitem{PRL}
I. Galanakis, G. Bihlmayer, V. Bellini, N. Papanikolaou, R.
Zeller, S. Bl\"ugel, and P.H. Dederichs, cond-mat/0105207.

\bibitem{valerio}
V. Bellini, N. Papanikolaou, R. Zeller, and P.H. Dederichs, {\em
Phys. Rev. B} {\bf 64} (2001) 094403.

\bibitem{Vosko}
S.H. Vosko, L. Wilk, and N. Nusair, {\em Can. J. Phys} {\bf 58}
(1980) 1200.

\bibitem{Zeller95}
R. Zeller, P.H. Dederichs, B. \'Ujfalussy, L. Szunyogh, and P.
Weinberger, {\em Phys. Rev. B} {\bf 52} (1995) 8807; N.
Papanikolaou, R. Zeller, and P.H. Dederichs, {\em J. Phys.: Cond.
Matter}, to be published.

\bibitem{Andersen86}
O.K. Andersen, Z. Pawlowska, and O. Jepsen, {\em Phys. Rev. B}
{\bf 34} (1986) 5253.

\bibitem{Stefanou}
N. Stefanou, H. Akai, and R. Zeller, {\em Comp. Phys. Commun.}
{\bf 60} (1990) 231.

\bibitem{Zeller97}
R. Zeller, {\em Phys. Rev. B} {\bf 55} (1997) 9400;
 K. Wildberger, R. Zeller, and P.H. Dederichs,
 {\em Phys. Rev. B} {\bf 55} (1997) 10074.

\bibitem{Zeller82}
R. Zeller, J. Deutz, and P.H. Dederichs, {\em Sol. St. Comm.} {\bf
44} (1982) 993; K. Wildberger, P. Lang, R. Zeller, and P.H.
Dederichs, {\em Phys. Rev. B} {\bf 52} (1995) 11502.

\bibitem{monkhorst}
H.J. Monkhorst and J.D. Pack, {\em Phys. Rev. B} {\bf 13} (1976)
5188.

\bibitem{ashcroft}
N.W. Ashcroft and N. D. Mermin, Solid State Physics, Saunders
College Publishing, 1976.

\bibitem{mackenzie} J.K. Mackenzie, A.J.W. Moore, and J.F. Nicholas,
{\em J. Phys. Chem. Solids} {\bf 23} (1962) 185 .

\bibitem{Feibelman90}
P.J. Feibelman and D.R. Hamann, {\em Surf. Sci.} {\bf 234} (1990)
377.

\bibitem{Feibelman92}
P.J. Feibelman,  {\em Phys. Rev. B} {\bf 46} (1992) 2532.

\bibitem{Mansfield}
M. Mansfield and R.J. Needs,  {\em Phys. Rev. B} {\bf 43} (1991)
8829.

\bibitem{Wan}
J. Wan, Y.L. Fan, D.W. Gong, S.G. Shen, and X.Q. Fan,
{\em Modelling Simul. Mater. Sci. Eng.} {\bf 7} (1999) 189; {\em and
references therein}.

\bibitem{jellium}
N.D. Lang and W. Kohn, {\em Phys. Rev. B} {\bf 1} (1970) 4555;
J.P. Perdew and R. Monniert, {\em Phys. Rev. Lett.} {\bf 37}
(1976) 1286; R. Monniert and J.P. Perdew, {\em Phys. Rev. B} {\bf
17} (1978) 2595; Z.Y. Zhang, D.C. Langreth, and J.P. Perdew, {\em
Phys. Rev. B} {\bf 41} (1990) 5674; J.P. Perdew, H.Q. Tran, and E.
Smith, {\em Phys. Rev. B} {\bf 42} (1990) 11627; K.F.
Wojciechowski, {\em Surf. Sci.} {\bf 437} (1999) 285.

\bibitem{schochlin}
J. Sch\"ochlin, K.. Bohnen, and K.M. Ho, {\em Surf. Sci.} {\bf
324} (1995) 113.

\bibitem{stumpf}
R. Stumpf and M. Scheffler,   {\em Phys. Rev. B} {\bf 53} (1996)
4958.

\bibitem{rodach}
Th. Rodach, K.P. Bohnen, and K.M. Ho,
 {\em Surf. Sci.} {\bf 286} (1993) 66.

\bibitem{Polatoglou}
H.M. Polatoglou, M. Methfessel, and M. Scheffler,  {\em Phys. Rev.
B} {\bf 48} (1993) 1877.

\bibitem{bross}
H. Bross and M. Kauzmann,  {\em Phys. Rev. B} {\bf 51} (1995)
17135.

\bibitem{eichler}
A. Eichler, J. Hafner, J. Furthm\"uller, and G. Kresse, {\em Surf.
Sci.} {\bf 346} (1996) 300.

\bibitem{morrison} L. Morrison, D.M. Bylander, and L. Kleinman,
{\em Phys. Rev. Lett.} {\bf 71} (1993) 1083.

\bibitem{weinert} M. Weinert, R.E. Watson, J.W. Davenport, and G.W. Fernando,
{\em Phys. Rev. B} {\bf 39} (1989) 12585.

\bibitem{wachter}
A. Wachter, K.P. Bohnen, and K.M. Ho, {\em Surf. Sci.} {\bf 346}
(1996) 127.

\bibitem{freeman}
H. Erschbaumer, A.J. Freeman, C.L. Fu, and R. Podloucky, {\em
Surf. Sci.} {\bf 243} (1991) 317.

\bibitem{takeuchi2}
N. Takeuchi, C.T. Chan, and K.M. Ho, {\em Phys. Rev. B} {\bf 43}
(1991) 14363.

\bibitem{needs}
R.J. Needs and M. Mansfield, {\em J. Phys.: Condens. Matter} {\bf
1}, (1989) 7555.

\bibitem{takeuchi}
N. Takeuchi, C.T. Chan, and K.M. Ho, {\em Phys. Rev. B} {\bf 43}
(1991) 13899.

\bibitem{eibler}
R. Eibler, H. Erschbaumer, C. Temnitschka, R. Podloucky, and A.J.
Freeman,  {\em Surf. Sci.} {\bf 280} (1993) 398.

\bibitem{Ho}
K.M. Ho and K.P. Bohnen, {\em Phys. Rev. Lett.} {\bf 59} (1987)
1833.

\end{thebibliography}
\end{document}